\documentclass[aps,preprint,preprintnumbers,showpacs,superscriptaddress]{revtex4-1}

\usepackage{graphics}
\usepackage{graphicx}
\usepackage{epsf} 
\usepackage{dcolumn}
\usepackage{bm}
\usepackage{amssymb}
\usepackage{amsmath}
\usepackage{amsbsy}
\usepackage{srcltx}
\usepackage{epsf}
\usepackage{graphicx}

\usepackage{graphicx,color,ifthen}

\newcommand{\rt}{{\mathcal T}}

\begin{document}


\title{Solid flow drives surface nanopatterning by ion-beam irradiation} 

\author{M.\ Castro}
\email[]{Corresponding author: marioc@upcomillas.es}
\affiliation{Grupo Interdisciplinar de Sistemas Complejos (GISC) and Grupo de Din\'amica No Lineal (DNL), Escuela T\'ecnica
Superior de Ingenier{\'\i}a (ICAI), \\ Universidad Pontificia Comillas, E-28015
Madrid, Spain}
\author{R.\ Gago}
\affiliation{Instituto de Ciencia de Materiales de Madrid, Consejo Superior de Investigaciones \\Cient\'{\i}ficas, E-28049 Madrid, Spain}
\author{L.\ V\'azquez}
\affiliation{Instituto de Ciencia de Materiales de Madrid, Consejo Superior de Investigaciones \\Cient\'{\i}ficas, E-28049 Madrid, Spain}
\author{J.\ Mu\~noz-Garc\'{\i}a}
\affiliation{Departamento de Matem\'aticas and GISC, Universidad Carlos III de Madrid,\\ Avenida de la Universidad 30, E-28911 Legan\'es, Spain}
\author{R.\ Cuerno}
\affiliation{Departamento de Matem\'aticas and GISC, Universidad Carlos III de Madrid,\\ Avenida de la Universidad 30, E-28911 Legan\'es, Spain}

\date{\today}


\begin{abstract}
Ion Beam Sputtering (IBS) is known to produce surface nanopatterns over macroscopic areas on a wide range of materials. However, in spite of the technological potential of this route to nanostructuring, the physical process by which these surfaces self-organize remains poorly understood. We have performed detailed experiments of IBS on Si substrates that validate dynamical and morphological predictions from a hydrodynamic description of the phenomenon. Our results elucidate flow of a nanoscopically thin and highly viscous surface layer, driven by the stress created by the ion-beam, as a description of the system. This type of slow relaxation is akin to flow of macroscopic solids like glaciers or lead pipes, that is driven by defect dynamics.
\end{abstract}

\maketitle
\medskip
\section{Introduction}

In the paper that historically coined the word ``Nanotechnology'' \cite{taniguchi:1974}, erosion of solid targets through ion-beam sputtering (IBS) was already put forward as the most promising technique 
to structure the surface of a wide range of materials. Indeed, for ion energies ranging from 100 eV to 100 keV, the procedure has shown a remarkable capability \cite{facsko:1999} to produce ordered nano-scale sized patterns (mostly ripples and dots) over large areas (up to tens of cm$^2$) for a wide range of targets, including semiconductors, metals, and insulators \cite{chan:2007,munoz-garcia:2009}. However, despite its large potential for technological applications \cite{smirnov:2003}, the promise of IBS as a fully controlled method to tailor patterns with custom designed properties has turned elusive. This is not only due to the challenges that description of the dynamics of surfaces in times of order 1 s pose to multi-scale modeling \cite{norris:2011} when microscopic events take place in times of order 1 ps. Rather, it originates in the lack of a basic understanding of the physical nature of the mechanism that controls the pattern formation.


Classically, since the seminal work by Bradley and Harper (BH) \cite{bradley:1988} the interplay between sputtering and surface diffusion had been identified as the key mechanism leading to pattern formation in IBS. Thus, a characteristic length scale would be selected \cite{cross_book} from the competition between the morphologically unstable dependence of
the sputtering yield with local surface curvature, and thermal surface diffusion that smooths out surface features
\cite{mullins:1957}. However, thus far only partial qualitative agreement had been reached between this classical description 
and experiments \cite{chan:2007}. Only after the recent realization of the non-trivial role of impurities in the emergence of the pattern \cite{ozaydin:2005,bradley:2010} has the need arisen to dispose of the BH mechanism for the simplest case of monoelemental semiconductor targets like silicon, that are amorphized by the ion beam \cite{gnaser:1999}, as in recent experiments on Ar$^+$ irradiation \cite{madi2008multiple,macko:2010}. For instance, one of the most direct implications of the BH picture, that ripple formation should take place for any incidence angle $\theta$ between the ion beam and the normal to the uneroded target, has been experimentally disproved, there being a critical angle $\theta_c$ such that ripples appear only for $\theta > \theta_c$. Incidentally, for ion energies between 10 and 40 keV, higher than in \cite{madi2008multiple,macko:2010}, this fact was already noted by Carter and Vishnyakov (CV) \cite{carter:1996}, who explained it phenomenologically via a smoothing effect of momentum transfer from the ions to the target atoms.

To date, two main non-BH-type mechanisms have been proposed to explain formation or absence of patterns in IBS of monoelemental semiconductors \cite{cuerno:2011}. Namely, mass redistribution \cite{moseler:2005,madi2011mass,norris:2011,hossain:2011} and ion induced solid flow \cite{surfsci2012}. The former employs results from Molecular Dynamics (MD) simulations in order to rephrase the CV effect as the influence of material (rather than momentum) displacement due to the beam, on the surface morphology. The latter proceeds, rather, through a continuum description of the surface flow that is driven by the surface confined stress due to the accumulated damage produced by irradiation. Remarkably, both descriptions agree with experiments with respect to e.g.\ the value of $\theta_c$ and the dependence of the pattern wavelength on $\theta$ \cite{cuerno:2011}. Note that, as far as physical mechanisms are concerned, MD simulations \cite{moseler:2005,madi2011mass,norris:2011,hossain:2011} are fundamentally limited in the IBS context, since they cannot probe the required macroscopic time scales at which non-trivial surface evolution takes place. This limitation has required e.g.\ the {\em ad-hoc} use of surface confined viscous flow \cite{umbach:2001} in \cite{madi2011mass,norris:2011}, in order to obtain the correct time scales for the description of the dynamics. What remains to be done is, rather, to prove if a physical description that incorporates viscous flow from the outset is indeed able to predict properties of the pattern that can be experimentally tested.

In this paper, we perform IBS experiments of Si in order to validate specific predictions from the hydrodynamic description put forward in \cite{surfsci2012}, that stem from the assumption that the main driving field in the process is the ion-induced residual stress that builds up in the flowing layer. Consistency between experiments and theoretical predictions thus allows us to elucidate the physical mechanism through which pattern formation occurs in these non-equilibrium nanoscopic systems.

\begin{figure}[!ht]
\begin{center}
\includegraphics[width=0.48\textwidth,clip=]{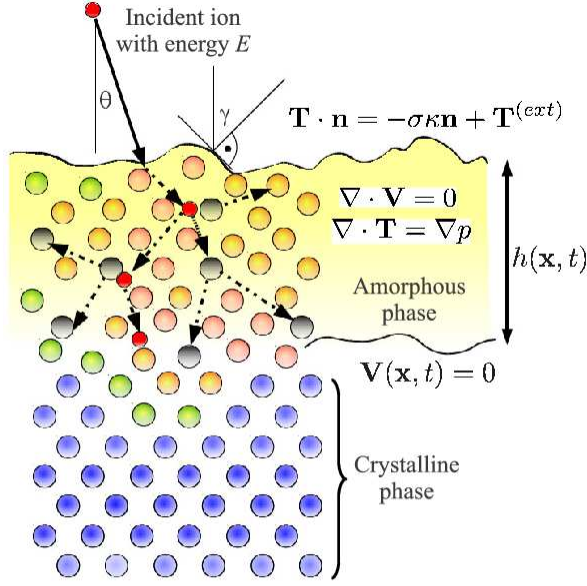}
\end{center}
\vspace{-3cm}
\caption{\label{cartoon} Schematic view of an IBS experiment. An incident
energetic ion (red) impinges onto the target at an angle $\theta$, inducing a
collision cascade that amorphizes a thin region through creation
of vacancies and interstitials. At long time scales, the amorphous solid
flows as a highly viscous fluid. Here $\gamma$ is the local slope of the
surface profile $h(x,t)$ (see the main text for notation and further details about the
equations).}
\end{figure}

\section{Predictions of the theory}
The main idea of the solid flow description of IBS \cite{surfsci2012} is that, as a consequence of the
impact of the ions and the subsequent release of energy within the target, defects are created inside the material. These events occur in a few picoseconds after the impact. Relaxation of some defects leads to sputtering of target atoms, but also to the generation of a residual stress that is confined to a thin amorphous layer that builds up beneath the surface \cite{kalyanasundaram:2006} and reaches a stationary thickness. This ion-induced (compressive) stress is characterized by a slow time relaxation which involves a highly viscous flow of the amorphous layer, that will be assumed to be incompressible, in a nanometric analogue of the motion of glaciers or lead pipes \cite{doake:1985}. The effect of these slow events can be cast into a hydrodynamical description \cite{oron:1997} of the form (see Fig.\ \ref{cartoon} for a schematic description of the system):
\begin{eqnarray}
\nabla\cdot{\bf V}=0, \quad \nabla\cdot {\bf T}&=&\nabla p, \label{Vp}\\
{\bf n}\cdot{\bf T}\cdot{\bf n}&=&-\sigma \kappa +T_n^{(ext)},\label{BCn}\\
{\bf n}\cdot{\bf T}\cdot{\bf t}&=&\nabla{\bf \sigma }\cdot{\bf t}+T_t^{(ext)},\label{BCt}\\
{\bf V}({\bf x},t)\left.\right|_{ac}&=&0,\label{noslip}
\end{eqnarray}
where ${\bf V}$ and $p$ are fluid velocity and pressure, ${\bf T}$ is the stress tensor inside the amorphous layer, $\bf n$ ($\bf t$) is the unitary vector locally normal (tangent) to the interface, $\sigma $ is the interface surface tension and $\kappa$ its curvature. Equations (\ref{BCn}) and (\ref{BCt}) are evaluated at the surface and Eq.\ (\ref{noslip}) at the amorphous-crystalline ($ac$) interface \cite{surfsci2012}. In our range of interest for ion energies (say, 300 to 1000 eV), the ripple wavelength (tens of nm) is much larger than the thickness of the amorphous layer (a few nm) \cite{chan:2007,munoz-garcia:2009}. Introducing this fact into the analysis of Eqs.\ \eqref{Vp}-\eqref{noslip} amounts to taking the so-called {\em shallow-water} limit \cite{oron:1997}. Using further a linear approximation in perturbations around a flat target profile, we obtain a real part for the linear dispersion relation (namely, the rate at which periodic perturbations with wave vector $q$ grow or decay \cite{cross_book}) that is given by
\begin{equation}
\omega _q=-\frac{f_Ed^3 \phi (\theta )}{3\mu }q^2-\frac{\sigma d^3}{3\mu }q^4,
\label{disprel}
\end{equation}
where $\phi (\theta )=\partial_\theta \left(\Psi(\theta )\sin(\theta )\right)$, $d$ is the average thickness of the amorphous layer, and $\mu$ is its (ion-induced) viscosity. The parameter $f_E$ can be understood as the gradient of residual stress induced by the ions across the amorphous layer, whose angular dependence is described through the function $\Psi(\theta )$. Since this angular function needs to be prescribed, we take $\Psi(\theta )=\cos\theta $ as the simplest geometrically motivated choice that shows good agreement \cite{surfsci2012} with previous experiments, as well as with the ones presented here.

Given $\Psi$ and $f_E$, we can extract testable predictions from the theory through the value of the ripple wavelength, $\lambda$, occuring to linear approximation of Eqs.\ \eqref{Vp}-\eqref{noslip},
\begin{equation}
\lambda =2\pi\sqrt{\frac{2\sigma }{-f_E\phi(\theta )}} .
\label{lambdac}
\end{equation}
The first prediction has to do with the angular dependence of the patterns. For $\Psi(\theta )=\cos\theta $ [hence, $\phi (\theta )=\cos(2\theta )$], we find a value for the critical angle $\theta_c=45^\circ$. 
Moreover, this value does not depend on the ion energy, in good agreement with X-ray experiments \cite{madi2011mass} and in contrast with an energy-dependent critical angle from MD simulations in \cite{norris:2011}, from $\theta_c \simeq 40^\circ$ at $250$ eV to $\theta_c \simeq 35^\circ$ at $100$ eV \footnote{This critical angle at $45^\circ$ can be also obtained \protect{\cite{surfsci2012}} without assuming a specific functional form for $\Psi(\theta)$ and assuming that, instead of a body force created by the ion, there is an elastic stress confined at the surface for an incompressible material through $T_t^{(ext)}$ in Eq.\ (\protect{\ref{BCt}}).}. Note, experiments at low energies ({\em e.g.}, $100$ eV) are difficult to perform reliably with a small dispersion.

Thus far, the analytical description is purely {\em kinematic}, in the sense that we have not made explicit which is the physical mechanism behind the stress. Experimentally, local measurements of stress in a nanoscopic amorphous layer
is problematic if not unfeasible. Thus, a natural procedure is to detail the nature of stress generation in the theoretical description, relating it with relevant physical parameters of the experiment, such as average ion energy, $E$, or flux, $J$, and obtain testable predictions from such an assumption. In \cite{davis1993simple} a simple (experimentally validated) model was proposed based on knock-on implantation and stress relaxation by defect migration to the surface. This leads to a dependence of the stress that builds up throughout the amorphous layer, $\rt\propto f_Ed$, on ion energy of the form
\begin{equation}
\rt(E)\propto \frac{Y}{1-\nu }\frac{E^{1/2}}{R/J + \tilde{\rho} E^{5/3}} .
\label{daviseq}
\end{equation}
Here, $Y$ and $\nu $ are the material Young's and Poisson's moduli, and $R$, $\tilde{\rho}$ are other material-dependent constants (related to properties like the binding energy). From Eq.\ (\ref{daviseq}) it follows that, for low energies, $\rt\sim JE^{1/2}$, while for large energies (which depend on $R$, but typically $E>100$ eV), $\rt\sim E^{-7/6}$.
We also need a prescription for the dependence of the amorphous layer thickness on energy. It is customary to assume a power law dependence of the form $d\sim E^{2m}$, where $m$ is assumed to be in the range $1/3-1/2$ and can be determined approximately from TRIM simulations \cite{ziberi:2005}. Thus, for a fixed angle, $f_E\propto \rt/d\sim E^{-7/6-2m}$, and using \eqref{lambdac} we find
\begin{equation}
\lambda\sim E^a,
\label{}
\end{equation}
with $a\in [0.92,1.08]$, namely, the dependence of the ripple wavelength with energy is roughly linear.

Another prediction can be obtained from Eq.\ (\ref{disprel}), that allows to find the characteristic scale for the exponential growth of the pattern amplitude occurring at short times during with linear approximation holds. Thus,
\begin{equation}
\tau(E,\theta ) \equiv 2\pi /\omega _{q=q^*} \sim \frac{\mu }{f_E^2d^3\phi^2  (\theta )}\sim \frac{E^b }{J\phi^{2} (\theta )},
\label{timescale}
\end{equation}
with $b \in [1.33,1.67]$, $q^*=2\pi /\lambda$, and we have used that $\mu \sim 1/J$ \cite{umbach:2001}.
Notice that e.g.\ the surface roughness, $W$, grows in the linear unstable regime as $W\propto e^{t/\tau}$. This exponential behavior will eventually be {\em interrupted} by nonlinear mechanisms (coming, for instance, from stress \cite{surfsci2012} or from purely erosive effects \cite{castro:2005}) at sufficiently long times. Hence, the time duration of the validity of the linear approximation is also controlled by $\tau $.

\section{Experiments}
Si(100) targets (380 m thick, p-type, $1-10$ Ohm cm) were sputtered with
Ar$^+$ ions under different incidence angles within the $300-1000$
eV energy range. The angle $\theta$ was adjusted by axial rotation of
the sample in front of the ion gun with an overall resolution of $\pm
2^\circ$. The ions were extracted from a commercial $3$ cm beam-diameter
Kaufman-type ion gun (VEECO) located $25$ cm away from the target. In order to
avoid metal contamination on the surface during irradiation, the sample holder
was covered with a sacrificial Si wafer and the sample was attached to it with
a double sided conductive vacuum tape. Prior to the process, the current density
at the sample position in the plane parallel to the source grids ($\theta =
0$) was set to $50$ $\mu $A/cm$^2$ with a Faraday cup located on a movable
shutter that prevents, additionally, unwanted irradiation of the sample.
The surface morphology of the irradiated surfaces was imaged ex-situ with a
Nanoscope IIIa equipment (Bruker) operating in intermittent contact mode and
using silicon cantilevers (Bruker) with a nominal radius of curvature of $10$
nm.

\begin{figure}[!ht]
\begin{center}
\begin{tabular}{cc}
\includegraphics[width=0.23\textwidth,clip=]{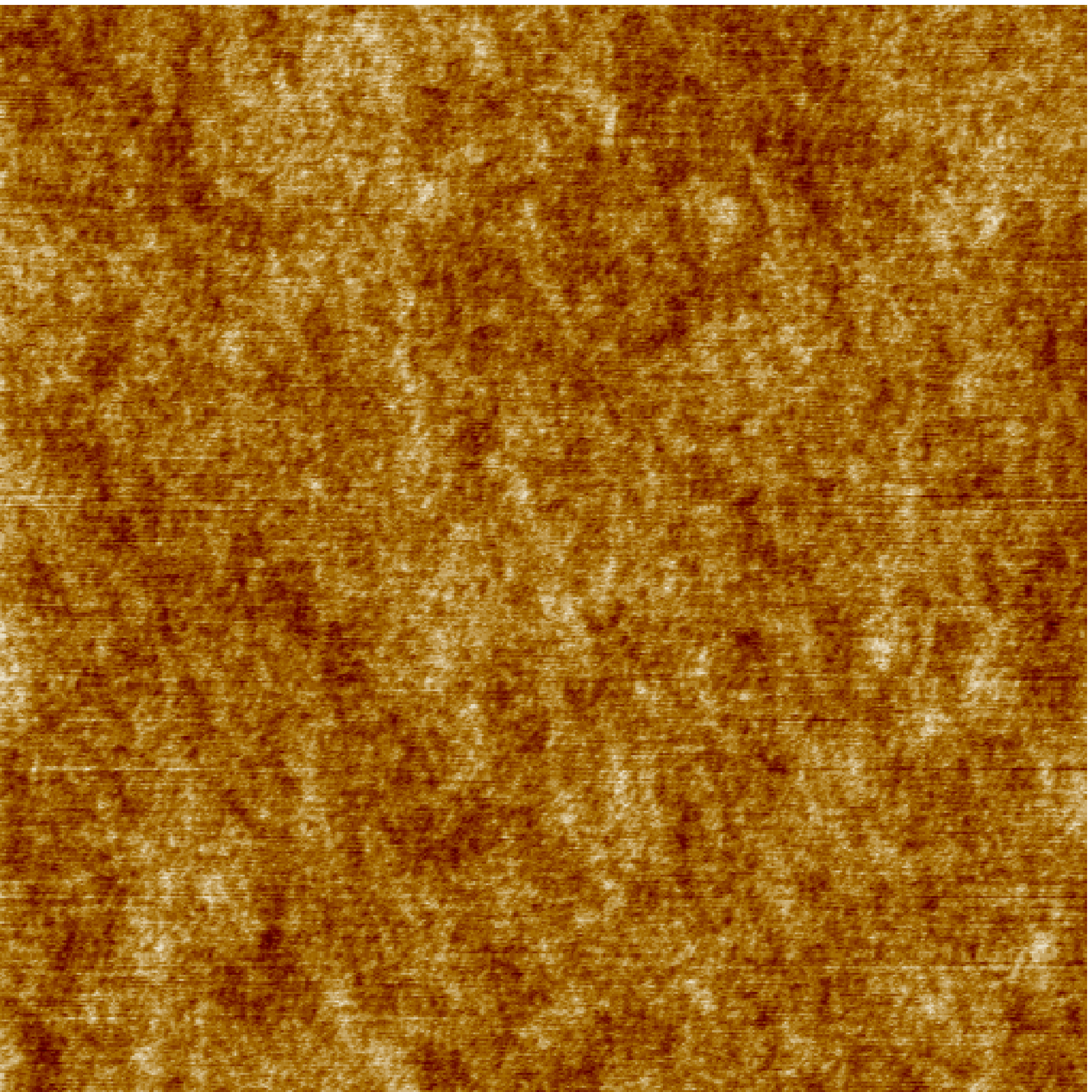} &
\includegraphics[width=0.23\textwidth,clip=]{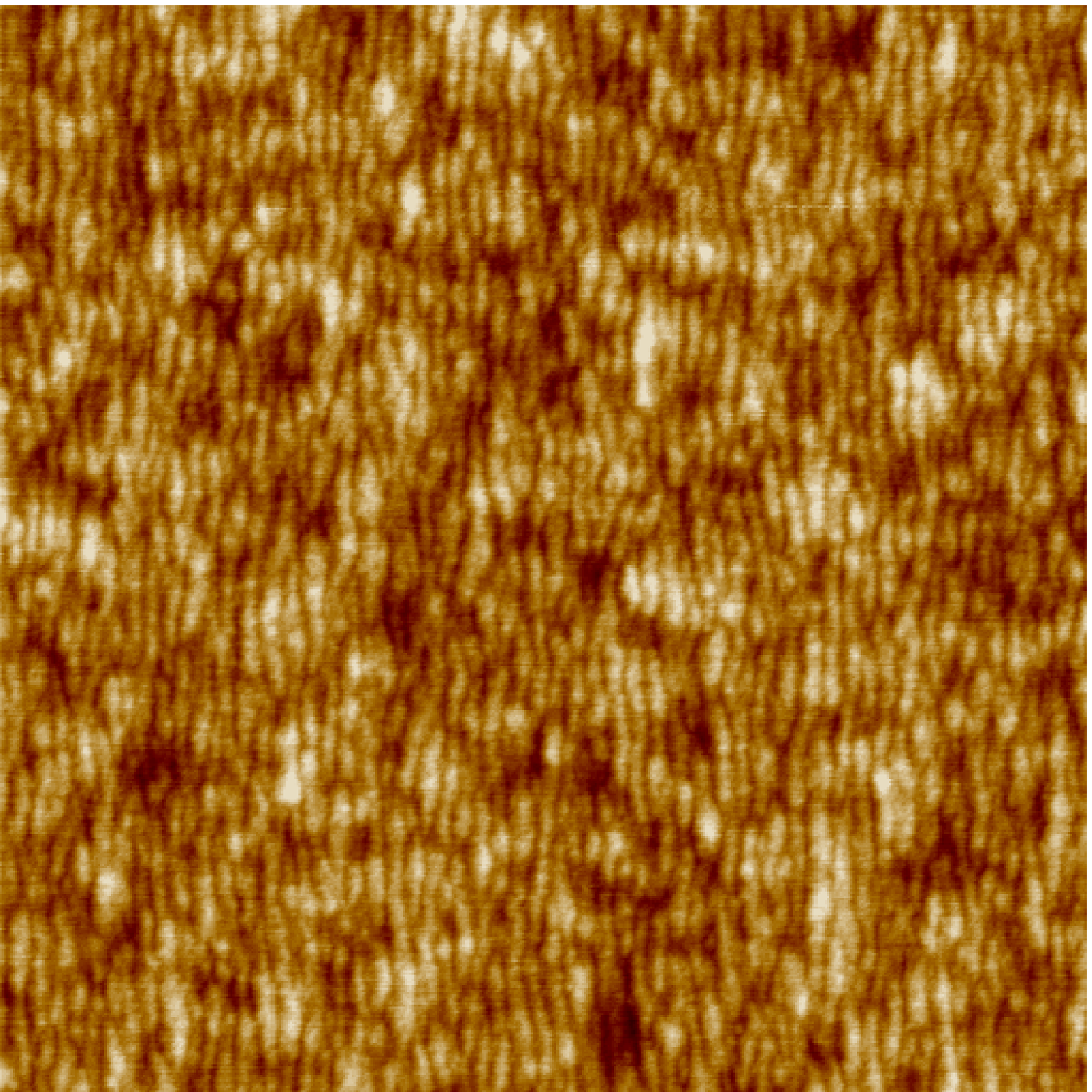} \\
(a)&(b) \\
\includegraphics[width=0.23\textwidth,clip=]{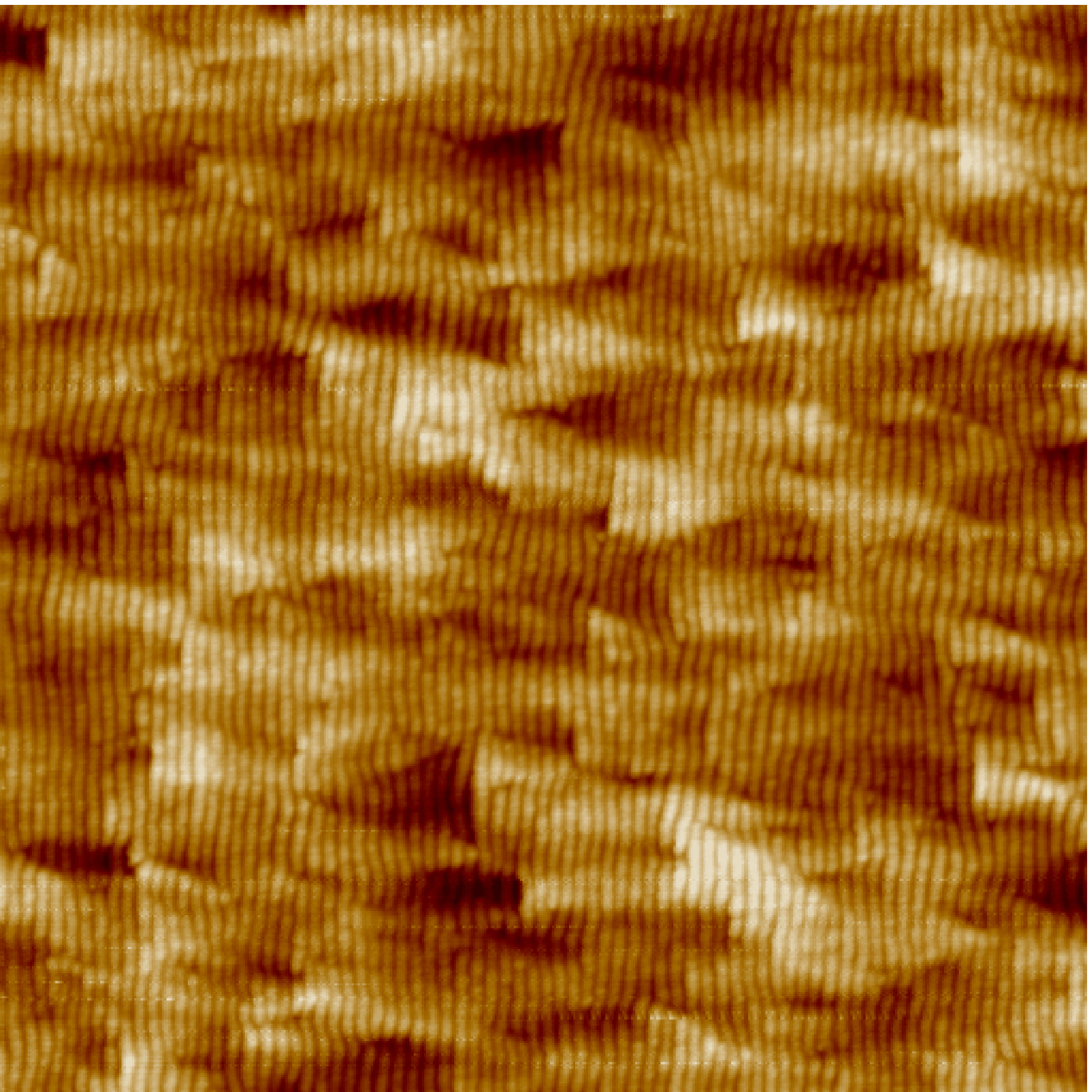} &
\includegraphics[width=0.23\textwidth,clip=]{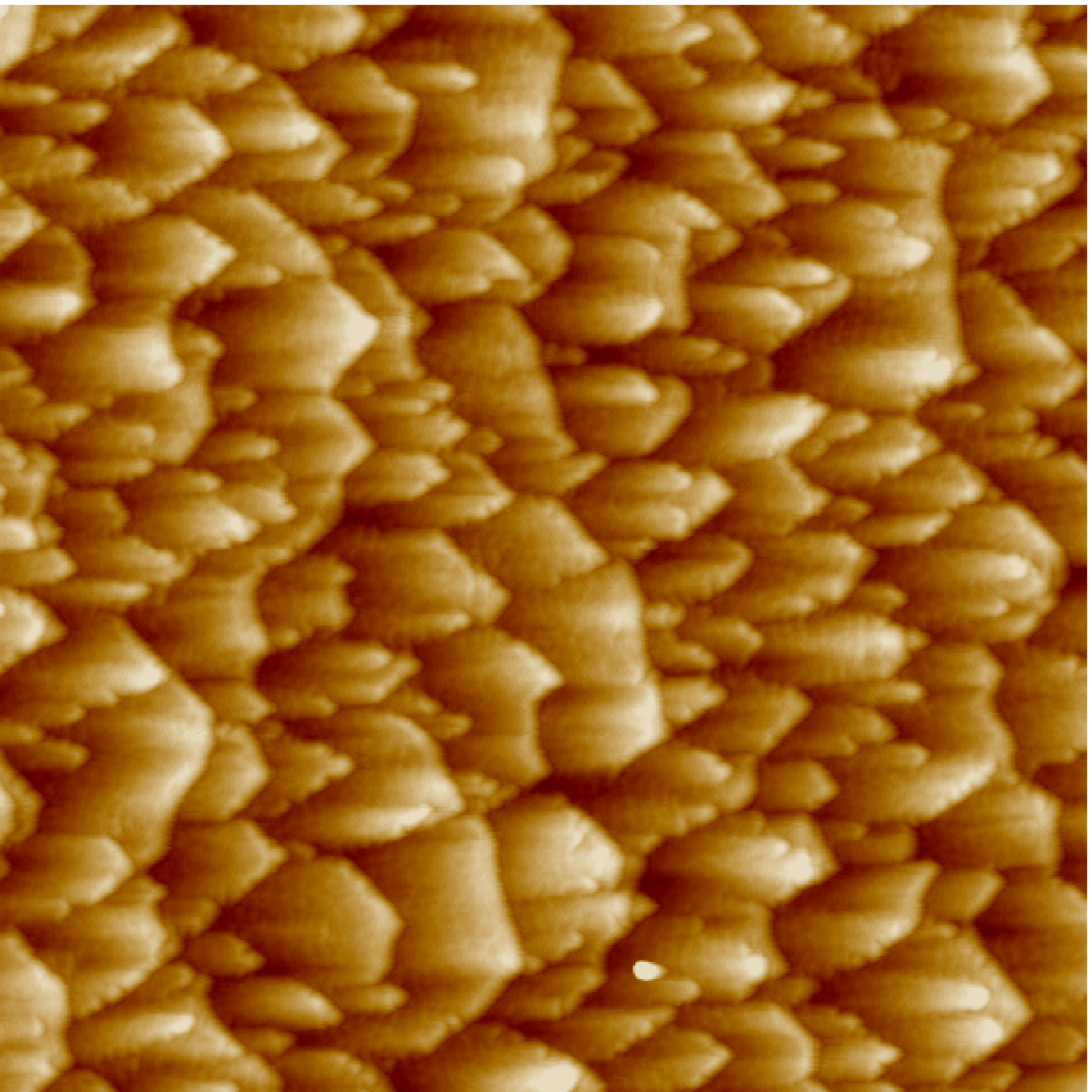} \\
(c)&(d) \\
\end{tabular}
\end{center}
\caption{\label{morphology} Experimental morphologies of ion irradiation of Ar$^+$ on Si. All the figures are obtained at constant dose, at energy $E=700$ eV, and at angles of incidence, $\theta=$ (a) $45^\circ$, (b) $55^\circ$, (c) $65^\circ$, (d) $75^\circ$. Note the absence of pattern at $45^\circ$. Lateral size is $2$ $\mu$m for all the panels, with ions arriving from right to left in all cases.
}
\end{figure}

\begin{figure}[!ht]
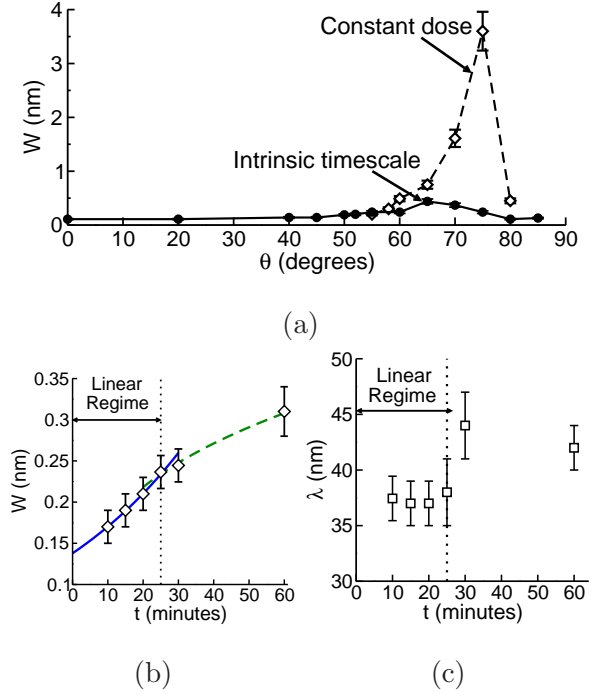

\begin{center}
\begin{tabular}{cc}
 \multicolumn{2}{c}{\includegraphics[width=0.45\textwidth,clip=]{W_angle} }\\
 \multicolumn{2}{c}{(a)}\\
 \includegraphics[width=0.23\textwidth,clip=]{Wtime}
 &\includegraphics[width=0.23\textwidth,clip=]{lambda_time} \\
(b)&(c)\\
\end{tabular}
\end{center}
\caption{\label{Wtheta} (a) Roughness dependence on incidence angle for experiments performed at constant dose (empty diamonds) and at the {\em intrinsic} time scale predicted by Eq.\ (\ref{timescale}) and estimated through (\ref{timescaling}) (solid black circles) for $E=700$ eV. (b) Time evolution of the surface roughness (circles) for $E=700$ eV and $\theta =55^\circ$. The blue solid line is an exponential fit (linear regime), while the green dashed line is a power law fit with exponent $0.32$. (c) Time evolution of the ripple wavelength (squares) for the same conditions as in (b). The time regime in which the wavelength grows with time (\!{\em coarsening}) starts at $t\simeq 25$ minutes as indicated by the dotted line; this is another signature of the onset of non-linear effects.  At shorter times the wavelength is almost constant. The same crossover time separates exponential from power law behavior in (b).}
\end{figure}

In Fig.\ \ref{morphology} we show four AFM micrographs of Si (100) surfaces bombarded at $E=700$ eV and different incidence angles. The surface does not contain any visible pattern for $\theta =45^\circ$ (or for smaller angles, not shown). On the other hand, as seen in Fig.\ \ref{Wtheta}a, roughness values increase significantly above this critical angle \cite{macko:2010}.

In order to compare experimental data with the model predictions derived above, we note that the linear regime
is expected to have a duration that changes with experimental conditions, see Eq.\ (\ref{timescale}). For instance,
it is expected to last shorter for increasing angles of incidence $\theta>\theta_c$ (in a form reminiscent of critical slowing down for continuous phase transitions \cite{cross_book,cuerno:2011}). Actually, Eq.\ (\ref{timescale}) allows us to control the experimental times, in order to guarantee that the system is truly evolving within linear regime, given a fixed reference experiment in which such state can be unambiguously assessed. Thus, given a pair of angle-energy reference values, $(\theta _{\rm ref},E_{\rm ref})$, we can extrapolate the value of $\tau$ for any other pair $(\theta ,E)$ through
\begin{equation}
\tau(\theta ,E)=\tau(\theta _{\rm ref},E_{\rm ref})\frac{J_{\rm exp}(\theta _{\rm ref},E_{\rm ref}) E_{\rm ref}^{-\frac{7}{3}+2 m} \phi^2( \theta_{\rm ref} )}{J_{\rm exp}(\theta ,E) E^{-\frac{7}{3}+2 m} \phi^2( \theta )} ,
\label{timescaling}
\end{equation}
where
$
J_{\rm exp}(\theta ,E)=J_{\rm exp}(E)\cos\theta 
$
is the flux used in a particular experiment at energy $E$ and angle $\theta $, with $J_{\rm exp}(E)$ being the flux at normal incidence. In our case, we choose $\theta _{\rm ref}\equiv 65^\circ$ and $E_{\rm ref}\equiv 700$ eV. Thus, from Eq.\ (\ref{timescaling}) we can extract the experimental times for different angles or energies which we define as {\em intrinsic} time scale, namely, the time at which experiments at different angles and/or energies are within linear regime.

In Fig.\ \ref{Wtheta} we show how the surface roughness depends on the conditions under which the experiments have been performed. Thus, we emphasize that doing the experiments at constant time (or fluence) may produce surfaces which are described by different regimes (linear or non-linear), depending on the value of $\theta$. The morphologies in Fig.\ \ref{morphology} correspond to the times obtained at constant dose in Fig.\ \ref{Wtheta} where typical nonlinear motifs, like facets, are recognized, specially in Figs.\ \ref{morphology}c-d. Note how the onset of these effects correlates for different observables \cite{munoz-garcia:2009} (power law growth of $W$, coarsening of $\lambda$) in Figs.\ \ref{Wtheta}b,c.
\begin{figure}[!ht]
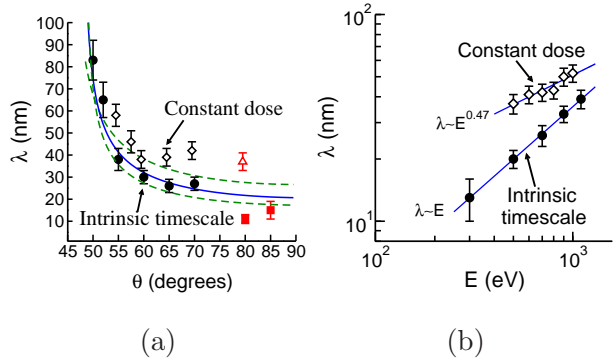

\begin{center}
\begin{tabular}{cc}
\includegraphics[width=0.24\textwidth,clip=]{lambda_angle}&
\includegraphics[width=0.24\textwidth,clip=]{lambda_energy}\\
(a)&(b)\\
\end{tabular}
\vspace{-5mm}
\end{center}
\caption{\label{lambdaangle} (a) Ripple wavelength dependence on the angle of incidence for $E=700$ eV. The empty diamonds correspond to {\em perpendicular} ripples like those in Fig.\ \ref{morphology}b, obtained at constant dose for all angles. The empty triangle is for {\em parallel} ripples with the same energy. Full symbols correspond to experiments performed in the proper linear regime as estimated by the {\em intrinsic} timescale in Eq.\ (\ref{timescale}). The solid blue line is a fit to Eq.\ (\ref{lambdac}), the green dashed lines corresponding to the same fit but assuming a $\pm 2^\circ$ uncertainty in $\theta$. The full black circles correspond to {\em perpendicular} ripples and the full red squares to {\em parallel} ripples. (b) Ripple wavelength dependence on the ion energy for constant dose experiments (empty diamonds) and intrinsic timescale experiments (solid black circles). The solid lines are fits to a power law with exponent $0.47$ and a linear dependence (as noted in the figure).
}
\end{figure}

In addition, in Fig.\ \ref{lambdaangle}a we show the dependence of the ripple wavelength on the incidence angle. The empty symbols stand for experiments performed at constant dose and the full ones are for experiments at the intrinsic timescale defined in Eq.\ (\ref{timescale}) that guarantees linear behavior. The blue solid line corresponds to a fit to Eq.\ (\ref{lambdac}) using $\Psi(\theta )=\cos\theta $, and the green dashed lines are a similar fit taking into account a $\pm 2^\circ$ uncertainty in the experimental measurement of $\theta$ \footnote{The fit can be further improved (not shown) if one also fits the critical angle  as $\lambda =2\pi\sqrt{\frac{2\sigma }{f_E\phi'(\theta_c )}}(\theta -\theta _c)^{-1/2}$}. These fits confirm the validity of Eq.\ (\ref{lambdac}). Moreover, in Fig.\ \ref{lambdaangle}b we show the dependence of the wavelength on the ion energy. Again, the prediction of the theory that $\lambda$ scales almost linearly with ion energy fits nicely, provided the experiments are performed at the intrinsic timescale, and not at (the customary) constant dose experiments. In general, agreement is not reached for the latter since those data points correspond to the nonlinear regime that sets in for times longer than the scale (\ref{timescale}), at which predictions made from linear approximation break down. This fact proves the self-consistency of our present analysis and stresses the predictive power of the solid flow theory.

\section{Discussion and conclusions}

We have presented theoretical predictions based on ion induced residual stress and viscous flow as the main physical mechanism driving surface nanopattern formation by IBS, together with the experimental validation of such predictions. These results allow us to extract important consequences. Note that the predictions in terms of ion energy originate from the physical model in \cite{davis1993simple}, in which direct knock-on implantation is assumed to produce stress that can be relaxed by defect migration to the free surface of the amorphous layer. When taken as an input of our hydrodynamic framework, this allows us to provide the {\em scaling} of different observables with energy. Thus, the ripple wavelength is predicted to scale roughly linearly with $E$, in marked contrast with the classical $E^{-1/2}$ scaling from BH theory \cite{bradley:1988,chan:2007,munoz-garcia:2009}. Other theoretical descriptions of the problem, both based on effective interface equations (see \cite{cuerno:2011} for an overview) or on atomistic simulations combined with multi-scale approximations, are based on mere superposition of mechanisms that actually operate at different time scales, rather than on a single framework that leads to a dynamic description as is the case here. As an example, in the original BH theory sputtering takes place in a few ps, while surface diffusion occurs in ns. One could actually expect contributions from sputtering to surface-confined transport \cite{umbach:2001,castro:2005}, but adding them together {\em ad-hoc}, as is natural to a first approximation, will necessarily overlook them.

Finally, an important part of our analysis relates to the dependence of the linear or nonlinear behavior with experimental parameters and observation time. Thus, an essential requirement for experimental reproducibility and meaningful comparison with theory is to make sure that measurements made for different conditions (e.g.\ energy, angle, etc.) correspond to the proper dynamical regime. We expect that clarification of the physical basis of IBS allows to enhance experimental control over the technique, that finally brings it up to the well-founded high expectations expressed almost 40 years ago \cite{taniguchi:1974}.

\begin{acknowledgments}
This work has been partially supported by grants FIS2009-12964-C05-01, FIS2009-12964-C05-03, FIS2009-12964-C05-04, and CSD2008-00023 (MICINN, Spain). J.\ M.-G.\ is supported by the Spanish MEC through the Juan de la Cierva program.
\end{acknowledgments}

\bibliographystyle{apsrev}
\bibliography{bib2}

\end{document}